\documentclass[twocolumn]{aastex631}

\usepackage{graphicx}
\usepackage{amsmath}
\usepackage{float}
\usepackage{natbib}
\usepackage{tabularx}
\usepackage{bm}
\usepackage{appendix}
\usepackage{longtable}
\usepackage{enumitem}

\newcommand{\msun}{$M_{\odot}$}
\newcommand{\zsun}{$Z_{\odot}$}
\newcommand{\emp}{EMP-Pathfinder }

\shorttitle{Binary Black Hole Mass Distribution}
\shortauthors{Ye et al.}

\graphicspath{{./}{figures/}}

\begin{document}

\title{Mass Distribution of Binary Black Hole Mergers from Young and Old Dense Star Clusters}%\footnote{Released on March, 1st, 2021}}

\author[0000-0001-9582-881X]{Claire S. Ye}
\affil{Canadian Institute for Theoretical Astrophysics, University of Toronto, 60 St. George Street, Toronto, ON M5S 3H8, Canada}
\correspondingauthor{Claire S.~Ye}
\email{claireshiye@cita.utoronto.ca}

\author[0000-0002-1980-5293]{Maya Fishbach}
\affil{Canadian Institute for Theoretical Astrophysics, University of Toronto, 60 St. George Street, Toronto, ON M5S 3H8, Canada}
\affil{David A. Dunlap Department of Astronomy and Astrophysics, University of Toronto, 50 St. George St., Toronto, ON M5S 3H4, Canada}
\affil{Department of Physics, University of Toronto, 60 St. George St., Toronto, ON M5S 3H8, Canada}

\author[0000-0002-4086-3180]{Kyle Kremer}
\affiliation{Department of Astronomy \& Astrophysics, University of California, San Diego, La Jolla, CA 92093, USA}

\author[0000-0002-8556-4280]{Marta Reina-Campos}
\affiliation{Canadian Institute for Theoretical Astrophysics, University of Toronto, 60 St. George Street, Toronto, ON M5S 3H8, Canada}
\affiliation{Department of Physics \& Astronomy, McMaster University, 1280 Main Street West, Hamilton, ON L8S 4M1, Canada}

\begin{abstract}
Dense star clusters are thought to contribute significantly to the merger rates of stellar-mass binary black holes (BBHs) detected by the LIGO-Virgo-KAGRA collaboration. We combine $N$-body dynamic models of realistic dense star clusters with cluster formation histories to estimate the merger rate distribution as a function of primary mass for merging BBHs formed in these environments. It has been argued that dense star clusters -- most notably old globular clusters -- predominantly produce BBH mergers with primary masses $M_p\approx30\,M_\odot$. We show that dense star clusters forming at lower redshifts -- and thus having higher metallicities -- naturally produce lower-mass BBH mergers. We find that cluster BBH mergers span a wide range of primary mass, from about $6\,$\msun\ to above 100\,\msun, with a peak near $8\,$\msun, reproducing the overall merger rate distribution inferred from gravitational wave detections. Our results show that most low-mass BBH mergers (about $95\%$ with $M_p\lesssim 20\,$\msun) originate in metal-rich ($Z \sim Z_{\odot}$) dense star clusters, while more massive BBH mergers form predominately in metal-poor globular clusters. We also discuss the role of hierarchical mergers in shaping the BBH mass distribution. Gravitational wave detection of dynamically-formed low-mass BBH mergers -- potentially identifiable by features such as isotropic spin distributions -- may serve as probes of cluster formation histories in metal-rich environments at low redshifts.
\end{abstract}

\section{Introduction} \label{sec:intro}

We are expecting to have more than 200 binary black hole (BBH) mergers detected by the LIGO-Virgo-KAGRA (LVK) collaboration from the fourth observing run later this year. This growing catalog of gravitational wave (GW) detections has tremendously expanded our understanding of massive star and binary evolution, as well as supernova mechanisms. One of the most fundamental properties of GW detections is the mass distribution of black holes (BHs). Since the first LVK detection \citep{GW150914}, BH masses have been used to distinguish between different BBH formation channels. Canonically, old and low-metallicity globular clusters have been associated with higher-mass BHs typically peaking around $20-30$\msun\,\citep{Rodriguez+2018_PNspin,Rodriguez+2018_PNeccentric,Rodriguez+2019_nextg,Mapelli+2022,Antonini+2023,Torniamenti+2024,Bruel+2024,Bruel+2025}. However, as the LVK catalog has grown, it has become clear that massive BHs are a subset of the full population, with the overall BH mass distribution peaking at roughly $8-10\,$\msun\, \citep{Abbott+2023_population,Callister_Farr_2024,Mali_Essick_2025,O4A}.  

How do these low-mass BBHs form? Studies have shown that isolated binary evolution in galactic fields could naturally account for BBH mergers with primary masses $\lesssim 50\,$\msun\, \citep[e.g.,][]{Giacobbo+2018,Mapelli_2021,vanSon+2022,vanSon+2023}. However, there is ongoing debate about whether dense star clusters (cluster mass $\gtrsim 10^5$~\msun\, and core density $\gtrsim 10^3\,M_{\odot}\,{\rm pc^{-3}}$; \citealp[see, e.g.,][]{PZ+2010}) can produce a significant number of BBH mergers with primary masses $\lesssim 20 M_{\odot}$ and where the primary mass distribution of cluster BBH mergers peaks. Understanding how the merger rate varies with BH mass can provide crucial insights into the formation pathways of BBH mergers, as well as the processes governing BH formation through massive star evolution and supernovae. \citet{Antonini+2023} used semi-analytical methods to study BBH mergers from globular clusters and suggested that the primary mass of merging BBHs peaks at around $30-40\,$\msun. Most recently, \citet{Bruel+2025} combined $N$-body dense star cluster simulations using the \texttt{Cluster Monte Carlo} code (\texttt{CMC}; \citealp{CMC1}) with the large-scale cosmological simulation \texttt{FIREbox} \citep{FIREbox} to probe massive star cluster formation across cosmic time and their contribution to merging BBHs. They showed that merging BBHs from dense star clusters have a merger rate peak at about $20\,$\msun. On the other hand, \citet{Torniamenti+2024} demonstrated that the pairing criterion for sampling BBH component masses in semi-analytical methods can significantly impact the mass distribution of BBH mergers. They compared two additional pairing criteria to the one in \citet{Antonini+2023} and showed that with a different pairing function they can produce many merging BBHs with primary mass $\lesssim 15\,$\msun, leading to a BBH merger rate that peaks at $\approx 10\,$\msun\, instead of at a higher mass. Furthermore, using \texttt{CMC} models, \citet{Chatterjee+2017_lowmassBBH} demonstrated that the BBH mass distribution is determined by the metallicity of their host clusters, where young, metal-rich clusters ($Z\gtrsim 0.5\,$\zsun) naturally produce low-mass merging BBHs, while old, metal-poor clusters tend to form BBHs with more massive components. Similar to isolated binaries in galactic disks, metallicity affects massive stellar winds. At high metallicity, stellar winds are stronger, leading to the formation of lower-mass BHs \citep[e.g.,][]{Belczynski+2010}. \citet{Banerjee_2022} also showed that young massive star clusters form merging BBHs whose merger rate peaks at around $10\,$\msun.

Dense star clusters' metallicity, mass, and formation rates at low redshift thus have significant impacts on the formation of merging BBHs with low-mass components. In addition, initial mass distributions and formation rates of dense star clusters determine the overall rates of dynamically-formed low-mass BBH mergers \citep{Rodriguez_Loeb_2018,Kremer+2020catalog,Bruel+2024}.

In this work, we study the mass distribution of BBH mergers by accounting for cluster formation histories as a function of redshift that follow star formation rates and comparing them to cluster formation simulations. We use a set of  \texttt{CMC} simulations designed to match closely the full range of properties of dense star clusters observed at present day, including both old globular clusters and young metal-rich clusters similar to those found in the Milky Way. This `backward modeling' approach is different from the `forward modeling' approach of \citet{Bruel+2025}, whose \texttt{CMC} cluster simulations employ initial properties derived from cosmological simulations and do not reproduce the massive and metal-rich clusters that are observed in the Milky Way \citep[][]{Grudic+2023,Rodriguez+2023}. We demonstrate that dense star clusters can produce in abundance a range of BH masses from around the lowest BH mass produced by supernovae to $\gtrsim 100\,$\msun\, produced by hierarchical mergers. Dense star clusters here refer to both young, metal-rich massive star clusters and old, metal-poor globular clusters. We introduce the properties of dense star clusters and the assumed cluster formation rates in Section~\ref{sec:clusters}. We describe the BBH merger rate as a function of mass and compare it to the rate observed by LVK in Section~\ref{sec:mass_distr}. We discuss uncertainties in Section~\ref{sec:discuss} and conclude in Section~\ref{sec:conclu}.

\section{Properties of Dense Star Clusters}\label{sec:clusters}

\begin{figure}
\begin{center}
\includegraphics[width=\columnwidth]{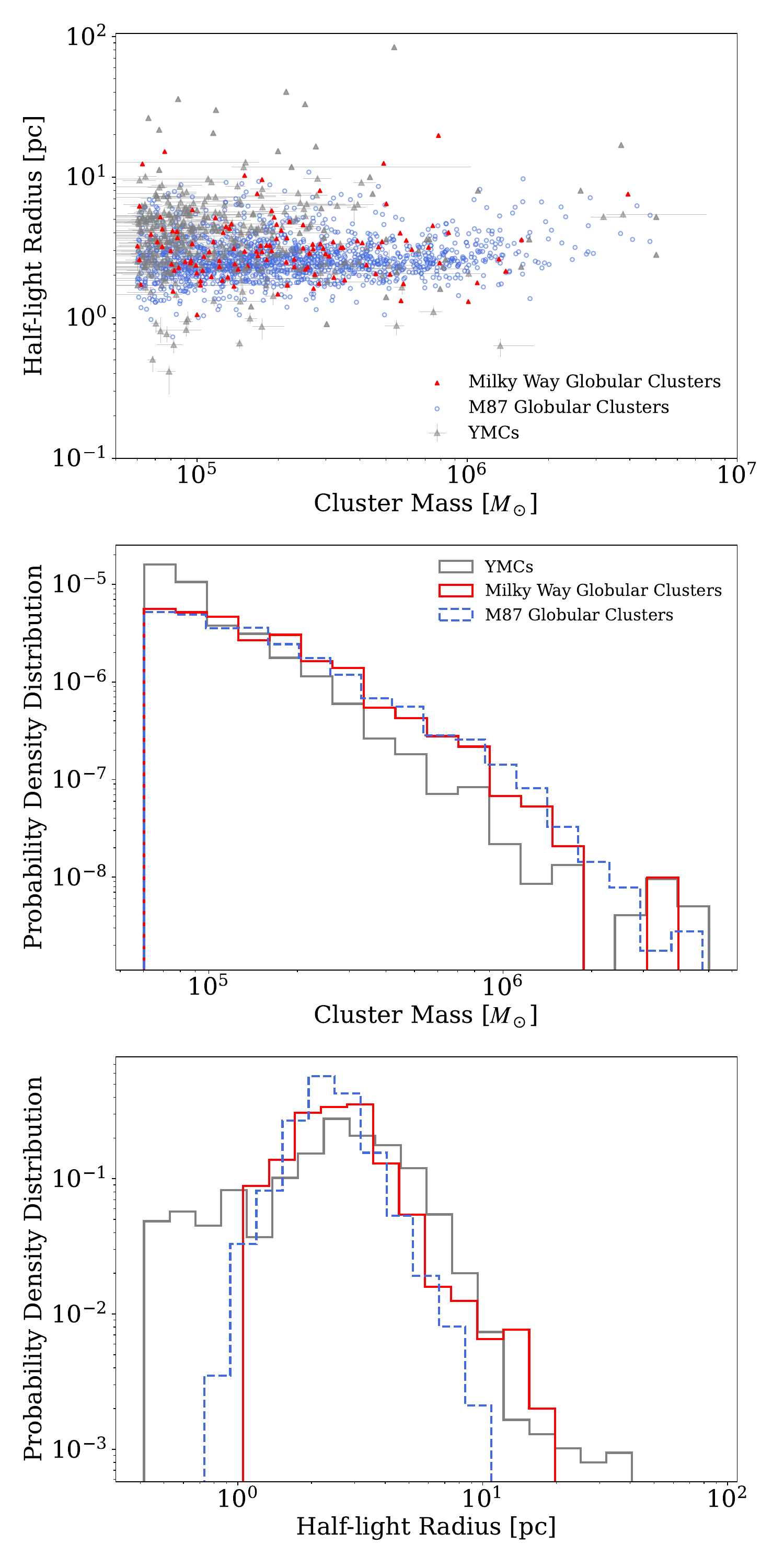}
\caption{Observed nearby young massive star clusters (YMCs) and globular clusters with mass $\gtrsim 6\times10^4\,$\msun\, for comparison. Milky Way globular cluster properties are taken from https://people.smp.uq.edu.au/HolgerBaumgardt/globular/. M87 globular clusters are from \citet{Jordan+2004}. We include YMCs with age smaller than 1~Gyr in 31 nearby galaxies from the Legacy Extragalactic UV Survey \citep[LEGUS;][]{Brown_Gnedin_2021} and from \citet{YMCs}.}\label{fig:YMCs}
\end{center}
\end{figure}

The formation and disruption of dense star clusters are closely connected to the star formation and evolutionary histories of their host galaxies \citep[e.g.,][]{Kruijssen_2025}, and in turn determine the mass and rate of dynamically-assembled BBH mergers. In particular, the formation and survival of young massive star clusters with high metallicities (loosely defined to be $\gtrsim 10^5\,$\msun\, and younger than a few Gyrs) directly affect the production of low-mass merging BBHs. A few young massive star clusters have been observed in the Milky Way, and hundreds have been identified in nearby galaxies. In Figure~\ref{fig:YMCs}, we show the masses and half-light radii of the observed nearby young massive star clusters from \citet{PZ+2010} and \citet{Brown_Gnedin_2021} and compare them to the observed Milky Way and M87 globular clusters. The overall mass and radius of these young and old dense clusters clusters are very similar, suggesting that the dynamical evolution of BHs in young massive star clusters--which depends on these cluster properties \citep[e.g.,][]{Rodriguez_Loeb_2018,Kremer+2020catalog}--is qualitatively similar to that of low-metallicity globular cluster counterparts.

It has been suggested that young massive star clusters may be the progenitors of systems similar to old and low-metallicity globular clusters \citep{PZ+2010}, making them essential in bridging small-scale star formation and large-scale galaxy formation. However, our understanding of star cluster formation and the evolution of their properties over cosmic time remains primitive due to the complex dependence of these processes on their evolving host galactic environments \citep[see, e.g.,][for a review]{Krumholz+2019}. JWST observations of proto-globular-cluster candidates at redshifts $z>2$ that are a few billion years old or less are providing unprecedented insights into these questions \citep{Mowla+2022,Vanzella+2022,Vanzella+2023,Adamo+2024,Rivera-Thorsen+2024,Senchyna+2024,Whitaker+2025}, but many uncertainties remain. At present, large-scale hydrodynamical cosmology simulations that couple the formation and evolution of massive star clusters with the evolution of galactic environments are the most sophisticated approach at probing the co-evolution of a population of galaxies and star clusters \citep[e.g.,][]{Pfeffer+2018,Reinacampos+2022,Grudic+2023}. Even so, it remains difficult to reproduce the observed Milky Way galaxy and its globular cluster system \citep[e.g.,][]{Kruijssen+2019,Rodriguez+2023}. Uncertainties in the evolution of the star formation rate, feedback, and stellar physics likely affect metal retention and the formation of metal-rich clusters in these simulations \citep{Grudic+2023}, which in turn significantly impact predictions for the properties of cluster BBH mergers. For example, the lack of massive and metal-rich star clusters in FIRE galaxy simulations \citep{Rodriguez+2023} likely have contributed to a deficit of low-mass BBH mergers from dense star clusters \citep[][their Figure~6]{Bruel+2025}.

\begin{figure}
\begin{center}
\includegraphics[width=\columnwidth]{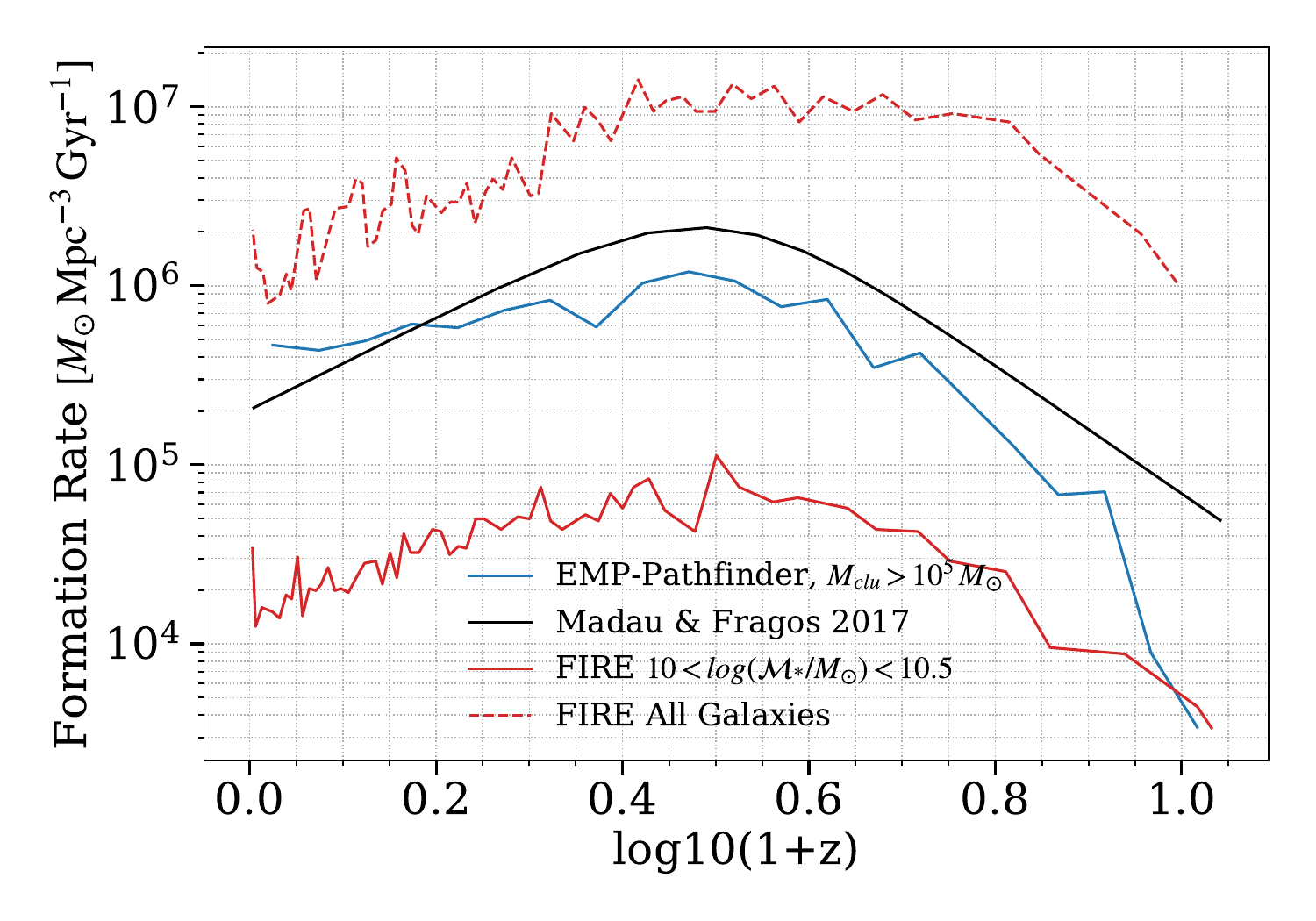}
\caption{Massive star cluster formation rate as a function of redshift. We assume the cluster formation rate follows the star formation rate in \citet{Madau_Fragos_2017} (black curve), normalized to a BBH merger rate of $36.2\,{\rm Gpc^{-3}\,yr^{-1}}$ at redshift $z=0.2$ for dense star clusters between about $10^5-10^6\,$\msun\, with an average initial cluster mass of about $3.6\times10^5\,$\msun. In comparison are the massive cluster formation rate from the \emp cosmological zoom-in Milky Way-mass simulations \citep{Reinacampos+2022} (blue curve) as well as massive cluster formation rates from \texttt{FIREbox} simulations \citep{Bruel+2025} (red curves). We scaled the rate from \emp to have the same unit as the other rates by assuming a galaxy density of $3\times10^6\,{\rm Gpc^{-3}}$, which is consistent with the observed massive galaxy densities \citep{Conselice+2016} and the simulated cluster formation rate from E-MOSAICS \citep{Joschko+2024}. The two FIRE curves show the rates for all galaxies and Milky Way-mass galaxies, respectively.}\label{fig:formation_rate}
\end{center}
\end{figure}

To maximize our understanding of the merger rates of dynamically formed BBHs--despite uncertainties in star formation and feedback--we explicitly couple testable cluster formation histories and properties with the dynamical evolution of BHs in dense star clusters. Here, we adopt a simple prescription which assumes that cluster formation follows star formation histories in \citet{Madau_Fragos_2017} and is described by 
\begin{equation}\label{eq:form_rate}
    \mathcal{R}_{\rm GC}(z) \propto \frac{(1+z)^{a_z}}{1+[(1+z)/(1+z_{peak})]^{a_z+b_z}}\,,
\end{equation}
where $a_z = 2.6$, $b_z=3.6$, and $z_{peak} = 2.2$ is the peak cluster formation redshift. This assumption matches well with the cluster formation histories from the latest cosmological zoom-in \emp simulations for the formation and evolution of massive star cluster populations in Milky Way-like galaxies \citep{Reinacampos+2022}. These simulations can reproduce the mass and metallicity distributions of observed Galactic globular clusters \citep[Figure~5 and Figure~9 in][]{Reinacampos+2022}. Figure~\ref{fig:formation_rate} compares the cluster formation rate that follows closely the redshift evolution of star formation from \citet{Madau_Fragos_2017} (black curve) to the cluster formation rate from \emp for star clusters with mass $>10^5\,$\msun\,(blue curve), assuming a galaxy density of $3\times10^6\,{\rm Gpc^{-3}}$.\footnote{Note that this number does not enter our calculation of BBH merger rates and mass distribution. It is only for normalizing the cluster formation rate from \emp \citep{Reinacampos+2022} for comparison with other star and cluster formation rates.} We normalize the \emp cluster formation rate so that its peak matches the formation rate of clusters with initial mass $>10^5\,$\msun\, from recent E-MOSAICS galaxy formation simulations for all galaxy masses across a $34^3\,{\rm Mpc^3}$ volume \citep[see Figure~6 in][]{Joschko+2024}. This normalization factor also agrees with observed galaxy densities for galaxies larger than $10^{10}\,$\msun\, at $z<2$ in \citet[][their Figure~7]{Conselice+2016}. The star and cluster formation rates both peak at around redshift $z=2$ and have very similar shapes throughout low and high redshifts. For comparison, we also plot the star cluster formation rates from the \texttt{FIREbox} cosmological simulation \citep[][their Figure~4, `SFRpeak' model]{Bruel+2025} for all galaxies (dotted red curve) and Milky Way-mass galaxies (solid red curve). All rates in Figure~\ref{fig:formation_rate} peak at redshift $\sim 2$, with small differences in the ascending and descending slopes and uncertainties in the absolute rates of cluster formation.

In addition to the cluster formation rate, we follow \citet{Fishbach_Fragione_2023} and \citet{Ye_Fishbach_2024} and adopt a Schechter function for the initial cluster mass distribution \citep{PZ+2010,Antonini_Gieles_2020} as described by
\begin{equation}
    p(M_{\rm GC}) \propto \left(\frac{M_{\rm GC}}{M_S}\right)^{\beta_S} {\rm exp} \left(-\frac{M_{\rm GC}}{M_S}\right)\,,
\end{equation}
where $\beta_S = -2$ and the Schechter mass $M_S = 10^{6.26}$\,\msun, a Gaussian distribution for the initial cluster virial radius distribution where
\begin{equation}
    p(r_v) \propto {\rm exp} \left(-\frac{(r_v - \mu_r)^2}{2\sigma_r^2} \right)\,,
\end{equation}
with mean $\mu_r=2~$pc and standard deviation $\sigma_r=2~$pc,
and a lognormal metallicity distribution $p(Z)$ at each formation redshift $z$ with $\sigma_Z = 0.5$ and $\mu_Z$ given by the redshift dependent equation from \citet{Madau_Fragos_2017},

\begin{equation}
    {\rm log}_{10} \langle Z/Z_{\odot} \rangle = 0.153-0.074z^{1.34}\,.
\end{equation}

We assume that these initial distributions are independent of redshift. Furthermore, we utilize the BBH mergers from a large catalog of 144 dense star cluster simulations \citep{Kremer+2020catalog} run with the state-of-the-art \texttt{CMC} Monte Carlo $N$-body code to study the mass distribution of merging BHs. The cluster simulations span a wide range of initial conditions in cluster mass ($1.2\times10^5$, $2.4\times10^5$, $4.8\times10^5$, and $9.7\times10^5\,$\msun), virial radius (0.5, 1, 2, and 4~pc), metallicity (0.0002, 0.002, and 0.02), and galactocentric distance (2, 8, and 20~kpc). 

We weight each cluster with the distributions of their initial mass $M_{\rm GC}$, virial radius $r_v$, metllicity $Z$, and formation histories from Equation~\ref{eq:form_rate}, and sum over all $n$ BBH mergers from a cluster to calculate the merger rate as follows
\begin{multline}
    \mathcal{R_{\rm BBH}}(z_l) = p(M_{\rm GC}^i) \Delta M_{\rm GC}^i p(r_v^j) \Delta r_v^j\\
    \times \sum_n \mathcal{R_{\rm GC}}(\hat{z}(t_l+t_m)) p(Z^k|\hat{z}(t_l+t_m)) \Delta Z^k\,.
\end{multline}
Here $t_l$ is the lookback time and $t_m$ is the BBH merger time since the formation of its host cluster. $\hat{z}$ is a cosmological function converting time to redshift. The total BBH merger rates from dense star clusters is estimated by summing over all \texttt{CMC} catalog models. For more details see \citet{Ye_Fishbach_2024}. 

Despite our semi-analytical approach compared to the FIRE simulations \citep[e.g.,][]{Rodriguez+2023,Bruel+2024,Bruel+2025}, which model cluster formation and evolution more self-consistently within the context of galaxy evolution, our method provides an upper limit on the formation of low-mass merging BBHs from dense star clusters by not suppressing the formation of metal-rich clusters at low redshift. Furthermore, we use cluster simulations that reproduce the properties of globular clusters in the Milky Way and the local Universe (see, e.g., Figure~\ref{fig:YMCs} and \citealp{Kremer+2020catalog}). This is crucial since observations of nearby dense star clusters offer the most robust constraints on massive cluster properties. This differs from \citet{Bruel+2025}, whose simulations do not reproduce the present-day mass distribution of observed Milky Way globular clusters \citep{Grudic+2023,Rodriguez+2023}.

\section{Mass Distribution of Black Holes}\label{sec:mass_distr}
\begin{figure*}
\begin{center}
\includegraphics[width=\textwidth]{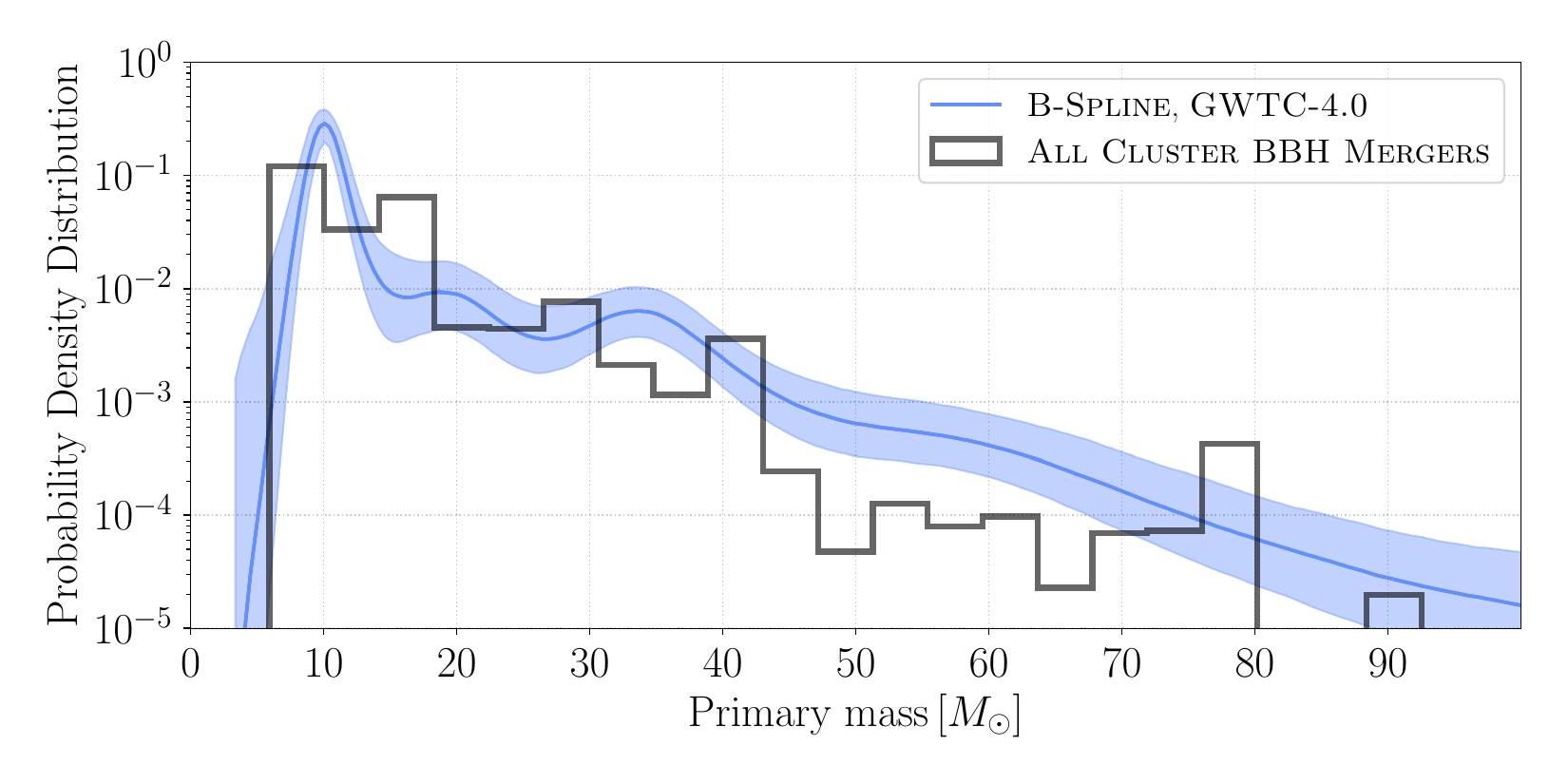}
\caption{Probability density distribution of BBH merger rates as a function of primary mass at redshift $z=0.2$. Black histogram shows BBH mergers from dense star clusters. The blue curves show BBH merger rates inferred non-parametrically from the latest Gravitational Wave Transient Catalog 4.0 \citep[GWTC-4.0;][]{O4A}, where the shaded area shows the 90\% credible interval.}\label{fig:mass_distri}
\end{center}
\end{figure*}

It is well established that dense star clusters can contribute significantly to the merger rates of BBHs \citep[e.g.,][]{Rodriguez_Loeb_2018,Antonini_Gieles_2020,Rodriguez_2021RNAAS,Banerjee_2022}. The distribution of BBH merger rates as a function of mass in the local Universe serves as an important indicator of their formation channels. We compare the BBH merger rate probability density as a function of primary mass from dense star cluster with those measured by GW detections in Figure~\ref{fig:mass_distri}. Dense star clusters, including young, metal-rich massive star clusters and old, metal-poor globular clusters, produce BBH mergers that match the overall mass distribution from GW detections. There is a peak at $\sim 8\,$\msun\, for cluster mergers, consistent with GW detections \citep{Abbott+2023_population,Callister_Farr_2024,Mali_Essick_2025,O4A}. We demonstrate, for the first time, that dense star clusters can reproduce the relative rates of low-mass versus high-mass BBH mergers. This result arises naturally when assuming that the cluster formation rates and metallicity follow those of the stars (Section~\ref{sec:clusters}). The more granular features of the distribution are sensitive to the uncertain stellar evolution models. In particular, the precise location of the low-mass peak depends on the detailed evolution of the progenitor massive stars and the supernovae mechanism\footnote{The \texttt{CMC} cluster simulations adopted the rapid supernova prescription from \citet{Fryer+2012}. A detailed exploration of different supernova prescriptions and their ability to reproduce GW observations is beyond the scope of this study and will be pursued in future work.} \citep[e.g.,][]{Fryer+2012,Sukhbold+2016,Burrows+2024} and may shift by a few \msun\ \citep[e.g.,][]{Sukhbold+2016}. The merger rate distribution turns over at $\sim 8\,$\msun\ and decreases at lower primary masses, where dynamically-formed BBH mergers can have primary masses as low as around $6\,$\msun.\footnote{Depending on supernova mechanisms and dynamical interactions, BHs in dense star clusters may have masses as low as about $3\,$\,\msun\, \citep[e.g.,][]{Ye+2024_lmgbh}.} The exact value of this lowest limit also depends on the physical processes mentioned above, including supernova natal kicks, as well as cluster properties which determine the mass segregation process and the dynamical evolution of BHs in dense star clusters \citep[e.g.,][]{Kremer+2020_bhrole,Ye+2020_dns,Ye_Fishbach_2024}. Overall, around $95\%$ of merging BBHs from dense star clusters are dynamically assembled, with similar fractions for systems with primary masses below or above $20\,$\msun.

\begin{figure}
\begin{center}
\includegraphics[width=\columnwidth]{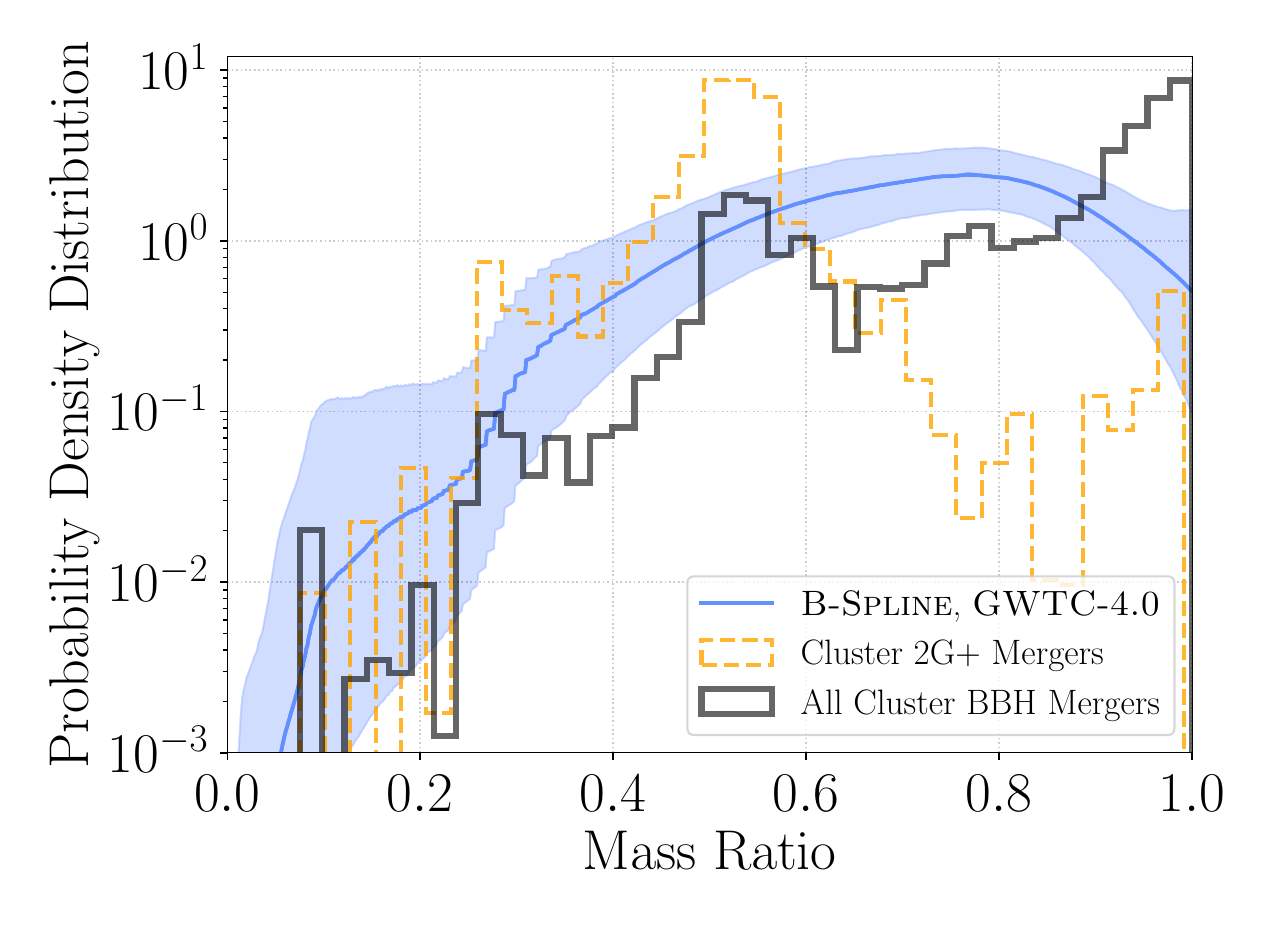}
\caption{Probability density distribution of BBH merger rates as a function of mass ratio at redshift $z=0.2$. The blue curves show the non-parametric model from GWTC-4.0 \citep{O4A}, where the shaded area shows the 90\% credible interval. Black histogram shows the mass ratio distribution of all BBH mergers from dense star clusters and the orange histogram shows only the hierarchical mergers. Higher-generation BHs preferentially merge with lower mass first-generation BHs in dense star clusters \citep[also see][]{Rodriguez+2019_nextg}.}\label{fig:mass_ratio}
\end{center}
\end{figure}

We also compare the mass ratio of BBH mergers from dense star clusters with GW measurements in Figure~\ref{fig:mass_ratio}. As expected from mass segregation, the mass ratio of all merging BBHs from clusters have a peak at around 1 for roughly equal mass mergers, and a secondary peak at around 0.5 from hierarchical mergers, consistent with previous studies \citep[e.g.,][]{Rodriguez+2019_nextg,Borchers+2025}. Most hierarchical mergers are between unequal-mass BHs \citep[also see Figure~4 in][]{Ye_Fishbach_2024}.

\subsection{Merger Rates as a Function of Primary Mass}\label{subsec:mass_z}
\begin{deluxetable}{c|c|c|c|c}
%\tabletypesize{\scriptsize}
\tablewidth{-1pt}
\setlength{\tabcolsep}{2pt}
\tablecaption{BBH Merger Rates at Various Primary Masses} \label{tab:rate_ratio}
\tablehead{\multicolumn{1}{c|}{Model} & \multicolumn{1}{c|}{$<20$~\msun} & \multicolumn{1}{c|}{$20-50$~\msun} & \multicolumn{1}{c|}{$>50$~\msun} & \colhead{Total}}
%\colhead{} & \colhead{\msun} & \colhead{} & \colhead{} & \colhead{}}
\startdata
Power law + Peak & $23.6^{+13.7}_{-9.0}$ & $4.5^{+1.7}_{-1.3}$ & $0.2^{+0.1}_{-0.1}$ & $28.3^{+13.9}_{-9.1}$ \\
Non-parametric & $25.0^{+79.9}_{-18.6}$ & $3.3^{+6.7}_{-2.2}$ & $0.2^{+0.4}_{-0.1}$ & $28.4^{+87.0}_{-20.9}$\\
B-SPLINE & $31.4_{-15.7}^{+31.5}$ & $4.4_{-2.0}^{+3.6}$ & $0.4_{-0.2}^{+0.6}$ & $36.2_{-17.9}^{+35.7}$\\
Cluster All & $32.7$ & $3.3$ & $0.2$ & $36.2$\\
Cluster 1G & $30.0$ & $2.0$ & $0.04$ & $32.04$\\
Cluster 2G+ & $2.7$ & $1.3$ & $0.13$ & $4.13$
\enddata
\tablecomments{BBH merger rates in $\rm{Gpc^{-3}\,yr^{-1}}$ at three ranges of primary mass and for all mergers. The Power law $+$ Peak rates are taken from the third observing run of LIGO/Virgo/KAGRA \citep{Abbott+2023_population}, while the non-parametric rates are calculated by integrating the merger rate distribution from \citet{Callister_Farr_2024} in each primary mass range. We also report the merger rates from GWTC-4.0, obtained by integrating the non-parametric B-SPLINE model over each primary mass range \citep{O4A}. All uncertainties shown are at the $90\%$ credible interval. We normalize the total cluster merger rate to be the same as the the total rate from the latest GW detections for simple comparison. This normalization is consistent with predicted rates of cluster BBH mergers \citep[e.g.,][]{Rodriguez_Loeb_2018,Kremer+2020catalog,Antonini_Gieles_2020}. The cluster merger rates are also sub-divided into rates of only first-generation mergers (`Cluster 1G') and for hierarchical mergers (`Cluster 2G+').}
\end{deluxetable}

\begin{figure*}
\begin{center}
\includegraphics[width=\textwidth]{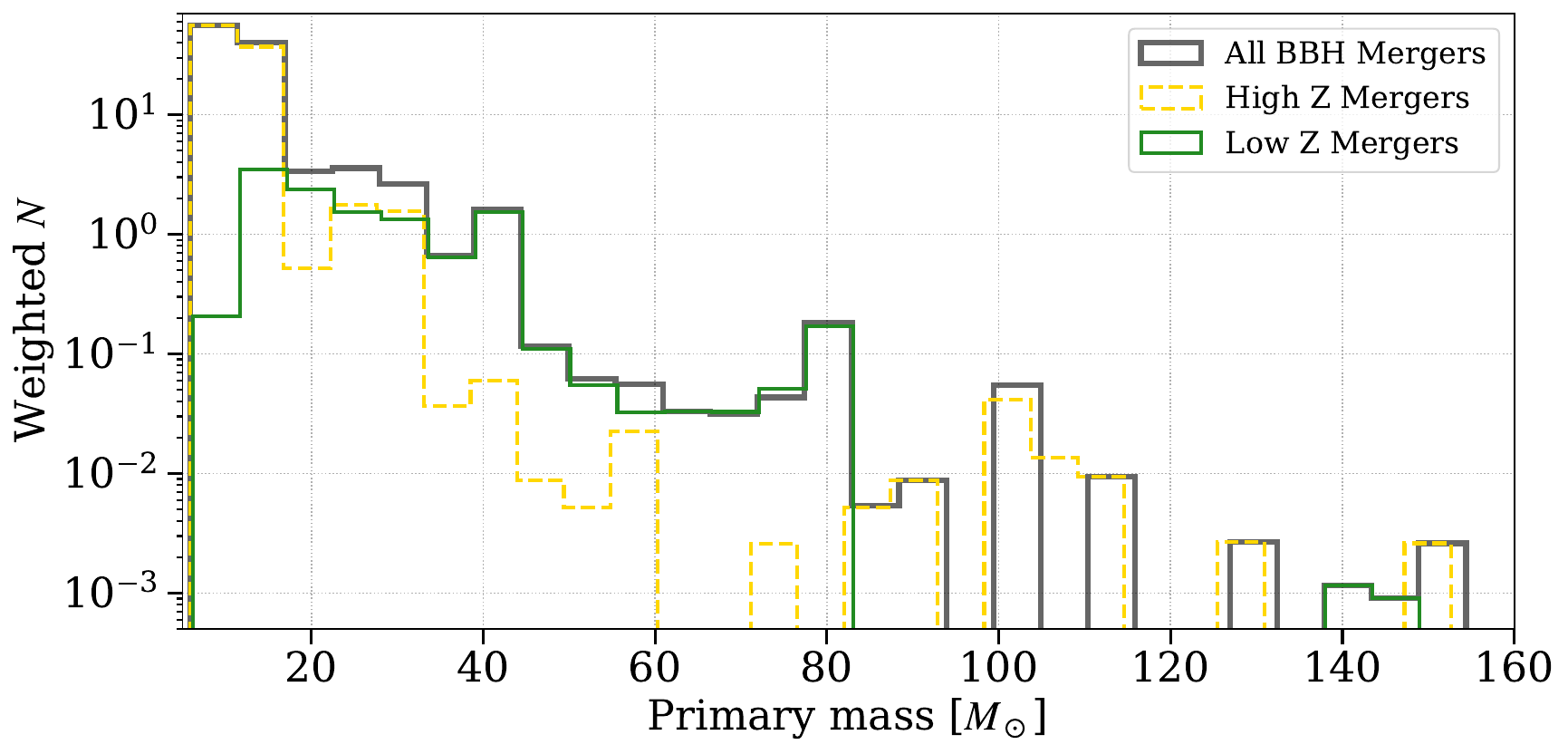}
\caption{Weighted probability distribution of BBH merger rates from dense star clusters at redshift $z=0.2$ for all BBH mergers and mergers from high- and low-metallicity clusters, respectively.}\label{fig:mass_metal}
\end{center}
\end{figure*}

We show the merger rates for three different primary mass ranges ($<20\,$\msun, $20-50\,$\msun, and $>50\,$\msun) as well as for all mergers in Table~\ref{tab:rate_ratio} and compare them to the Power law + Peak model from \citet{Abbott+2023_population} and the non-parametric model from \citet{Callister_Farr_2024}. GW detections show that there are roughly a few times more mergers with primary mass $M_p < 20\,$\msun\, than those with $20\,$\msun$< M_p <\,50$\msun; in turn, these are around an order to two of magnitude more abundant than mergers involving the most massive BHs with $M_p>50\,$\msun. The ratios of cluster BBH merger rates in the three primary mass ranges match closely with the GW detections.

Most BBH mergers from dense star clusters have primary mass $M_p<20\,$\msun. Specifically, most mergers with $15\,$\msun$\lesssim M_p<20\,$\msun\, (about $58\%$) are produced by metal-poor ($Z\lesssim10\%Z_{\odot}$) globular clusters, while most mergers with $M_p\lesssim 15\,$\msun\, (about $99\%$) are formed in metal-rich ($Z\approx\,Z_{\odot}$) massive star clusters. Figure~\ref{fig:mass_metal} shows the distribution of cluster BBH merger rates as a function of primary mass weighted by cluster formation histories for metal-poor and metal-rich clusters at the local Universe.

At the same time, metal-poor clusters are critical for producing merging BBHs with $M_p\gtrsim 20-30\,$\msun\, as metal-poor stars tend to leave behind more massive stellar remnants \citep[e.g.,][]{Belczynski+2010}. This agrees with previous studies of BBH mergers from dense star clusters \citep[see, e.g.,][]{Chatterjee+2017_lowmassBBH,Rodriguez+2019_nextg,Antonini+2023,Torniamenti+2024,Bruel+2024}. For the most massive BHs with $M_p\gtrsim100\,$\msun, however, metal-rich clusters dominate their merger rates at the local Universe (Figure~\ref{fig:mass_metal}). This may be because the most massive BHs have already merged at higher redshifts in metal-poor clusters \citep[e.g.,][]{Ye_Fishbach_2024} and are no longer detected in the local Universe. These massive BHs mainly acquire their mass from massive star progenitors formed via repeated stellar mergers in the most massive, metal-rich clusters.

\subsection{Hierarchical Mergers}\label{subsec:highg}
\begin{figure*}
\begin{center}
\includegraphics[width=\textwidth]{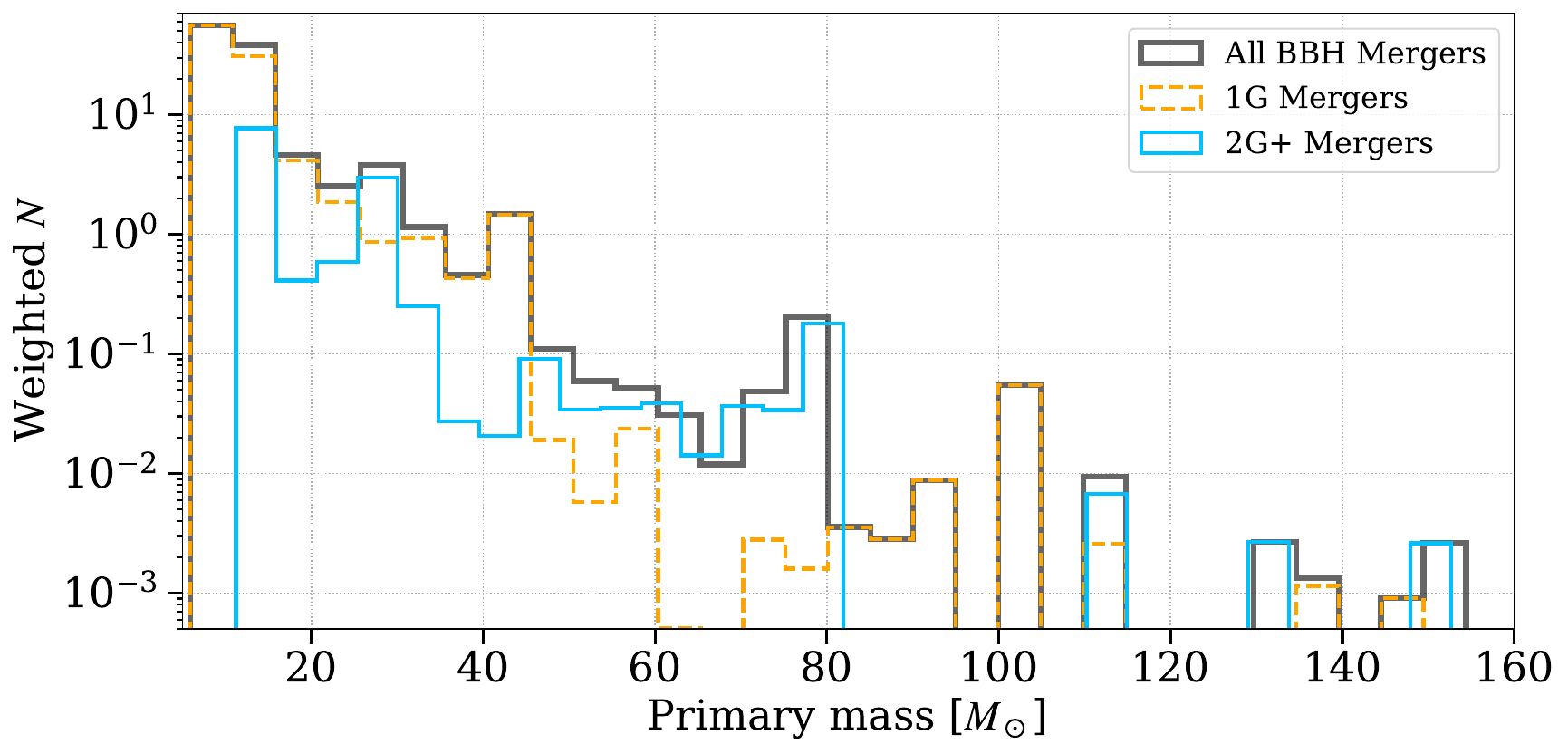}
\caption{Weighted probability distribution of BBH merger rates from dense star clusters at redshift $z=0.2$. We showed all BBH mergers (black histogram), mergers that contain only first-generation BHs (orange dashed histogram), and mergers that have higher-generation BHs (blue histogram).}\label{fig:mass_gen}
\end{center}
\end{figure*}

Repeated BH mergers also affect the mass distributions of merging BBHs. Here we define first-generation BHs as BHs formed directly from the collapse of massive stars and higher-generation BHs as BHs formed through previous BH mergers. About $90\%$ of cluster mergers involve only first-generation BHs and about $10\%$ contain BHs from previous mergers (Table~\ref{tab:rate_ratio}), consistent with previous study for BHs formed with no spin \citep{Rodriguez+2019_nextg}. These hierarchical mergers contribute to a large fraction of merging BBHs with $M_p\gtrsim30\,$\msun\, as shown in Figure~\ref{fig:mass_gen} and Table~\ref{tab:rate_ratio}. In particular, about $40\%$ of mergers with $M_p$ between 20~\msun\, and 50~\msun\, and almost all mergers with $M_p>50\,$\msun\, are hierarchical mergers. There are a few first-generation BBH mergers with $M_p>80\,$\msun\, in Figure~\ref{fig:mass_gen}. These massive BHs are produced by the collapse of massive stars formed in repeated stellar collisions \citep[e.g.,][]{Dicarlo+2019,Kremer+2020_umg,Gonzalez+2021}

In addition to affecting the mass distribution, repeated mergers leave an imprint on the spin of the higher-generation BHs determined by the orbital angular momentum of the merging binaries \citep[e.g.,][]{Buonanno2008}. Unlike first-generation BHs which are expected to be nearly non-spinning \citep[e.g.,][]{FullerMa2019}, higher-generation BHs retained in dense star clusters are expected to have spins spanning a broad range from about 0.4 to 1 \citep[e.g.,][]{GerosaBerti2017,Fishbach+2017_spin,Rodriguez+2019_nextg,Borchers+2025}. The spin distribution typically shows a major peak above 0.7, and potentially a smaller peak below 0.7, depending on the escape velocity of their host clusters, as well as the initial BH masses and spin configurations \citep{Borchers+2025}. Because of the differing spins of first- and higher-generation BHs, we expect the spin distribution of merging BBHs from dense star clusters to vary with their primary mass. For merging BBHs with $M_p\lesssim 20\,$\msun, where about $90\%$ of them contains only first-generation BHs, their component spin distribution is expected to be narrower and peak closer to zero (if the birth spin is zero) compared to merging BBHs with $M_p\gtrsim 20\,$\msun\, where a large fraction contains higher-generation, fast-spinning BHs \citep[see, e.g., Figure~13 in][]{Borchers+2025}. Nevertheless, it is interesting to note that most hierarchical mergers have primary masses below 20~\msun\,(Table~1).

\section{Summary and Uncertainties in Cluster Formation Histories}\label{sec:discuss}

In this study, we showed that the merger rate distribution of BBH mergers from dense star clusters spans a broad range of primary BH masses, from around $6\,$\msun\, to $\sim 100\,$\msun, with a peak at $\sim 8\,$\msun\, in the local Universe. We utilized BBH mergers from Monte-Carlo $N$-body simulations evolved over a Hubble time for dense star clusters spanning a range of initial masses, metallicities, virial radii, and galactocentric distances that reproduce the observed Milky Way globular clusters \citep{Kremer+2020catalog}. These simulations compute the detailed dynamical evolution of all stars coupled to stellar and binary evolution, providing realistic properties of BBH mergers including their component masses and merger times. We estimated the merger rate distribution of cluster BBHs in the local Universe by assuming that the redshift evolution of the cluster formation rate follows the star formation rate evolution from \citet{Madau_Fragos_2017}, and that the cluster initial mass function distribution follows a constant Schechter function with a slope $\beta_S=-2$ and an upper truncation of cluster mass at $M_S=10^{6.26}\,$\msun\ \citep{PZ+2010,Antonini_Gieles_2020}. These simple assumptions are consistent with more detailed studies of cluster formation which show that the cluster formation efficiency increases with star formation rate surface density (\citealp[e.g.,][]{Adamo+2020,Kruijssen_2025}; cf. \citealp{Krumholz+2019}) and that the initial cluster mass function follows a power law with a slope of about -2 across the mass range of dense star clusters \citep[For reviews, see][]{Krumholz+2019,Kruijssen_2025}.

However, the upper truncation of the cluster mass function likely does not remain constant; instead, it depends on properties such as gas pressure and star formation rate in star-forming environments, which are influenced by galaxy evolution and galaxy merger histories that likely vary over time \citep[e.g.,][]{Kravtsov_Gnedin_2005,Reinacampos+17,Adamo+2020,Grudic+2023,Bruel+2024}. For example, the formation histories of massive clusters may differ when environmental effects are taken into account \citep[e.g.,][their Figure~1]{Reina-Campos+2019}, and that the formation of massive clusters may be suppressed at low redshift. The upper truncation is critical since the number of dynamically-formed BBH mergers directly depends on the host cluster mass, with massive clusters dominating the merger rate \citep[e.g.,][]{Rodriguez+2016,Rodriguez_Loeb_2018,Bruel+2024}. Detailed cluster formation simulations with \emp show that an evolving initial Schechter mass function -- dependent on gas surface density, epicyclic frequency, and the Toomre parameter \citep{Toomre1964} -- produces present-day cluster mass distributions similar to those resulting from a constant power-law initial mass function with an upper mass truncation at $10^8\,$\msun\ \citep[][their Figure~8]{Reinacampos+2022}. Thus, our adoption of a constant initial cluster mass function provides a realistic approximation of detailed cluster formation simulations.

In addition, the stellar metallicity evolution across cosmic time directly affects the remnant mass of binary and stellar evolution. Up to a few percent of star formation may be in low-metallicity environments at low redshift past the peak of star formation \citep[e.g.,][]{Chruslinska_Nelemans_2019}. Metal-poor galaxies in the local group such as the Small Magellanic Cloud are also found to host young massive star clusters \citep[e.g.,][]{Dias+2010}. Our adopted lognormal distribution of metallicity as a function of redshift has a 0.5 dex spread in metallicity across all redshift, allowing for the formation of metal-poor clusters even at low redshift. This agrees with the spreads mentioned in \citet{Madau_Fragos_2017} and is only about a factor of 2 from the metallicity uncertainties discussed in \citet{Chruslinska_Nelemans_2019}. Uncertainties in the mean metallicity $\langle Z/Z_{\odot}\rangle$ across redshift will likely cause only minor variations in how rapidly the merger rate decreases with increasing primary mass, without significantly affecting the overall peak around $8\,$\msun, since dense star clusters with metallicities as low as $\sim 0.5\,Z_\odot$ can still produce these low-mass BBH mergers \citep{Chatterjee+2017_lowmassBBH}.

Because the number and properties of cluster BBH mergers depend on their host clusters’ mass and metallicity, the (non-)detection of low-mass BBH mergers from clusters by LVK would therefore provide valuable constraints on the role of environmental effects in cluster formation at low redshift. A significant depletion of dense star clusters with near-solar metallicity would result in a lack of dynamically-formed BBH mergers with primary masses $M_p\lesssim 20\,$\msun, as shown in \citet{Bruel+2025}. On the other hand, GW detections of low-mass merging BBHs with isotropic spin distributions—and possibly a fraction with large spins from hierarchical mergers—suggest that a non-negligible number of metal-rich, massive star clusters are producing merging BBHs.

Furthermore, BBH mergers are likely produced through multiple formation channels, including dense star clusters and isolated binaries in galactic fields \citep[e.g.,][]{Zevin+2021}. Lower-mass ($\lesssim 10^4\,$\msun), young or open star clusters with high metallicities could also produce many BBH mergers with primary mass $\lesssim 50\,$\msun\ \citep[e.g.,][]{Kumamoto+2020,DiCarlo+2020,Rastello+2021,Banerjee2021,Banerjee_2022}. We focused on dense star clusters with mass $\gtrsim 10^5\,$\msun\ in this study, as the more massive clusters likely produce the majority of dynamically formed BBH mergers \citep[e.g.,][]{Rodriguez_Loeb_2018,Antonini_Gieles_2020}. A simple estimate, assuming an initial cluster mass distribution scaling as $M^{-2}$ and a BBH merger yield scaling with cluster mass as $M^{1.6}$ \citep{Antonini_Gieles_2020}, suggests that lower-mass clusters (with initial masses of about $10^2-10^5\,$\msun) contribute only about $15\%$ of all merging BBHs formed in clusters (with initial masses spanning about $10^2-2\times10^6\,$\msun). This scaling relation for the merging BBH yield has also been verified within 1$\sigma$ uncertainties by \citet{Fishbach_Fragione_2023} and \citet{Mai+2025}, both of which use \texttt{CMC} simulations. In addition, \citet{Hong+2018}, using a different Monte Carlo $N$-body code \texttt{MOCCA}, found a scaling of around $M^{1.3}$ down to cluster masses of $\sim10^5\,$\msun, again consistent with the above relation at the 1$\sigma$ level. If lower-mass clusters follow a more linear scaling, as suggested by \citet{Banerjee_2018}, it would imply that massive, dense star clusters contribute an even larger fraction of BBH mergers across the full cluster mass range. If hierarchical mergers in dense star clusters are required to explain all merging BBHs with $M_p\gtrsim 50\,$\msun, there might be an overproduction of lower-mass BBH mergers from the combined star cluster and isolated binary channels. This suggests that 1) dense star clusters may produce fewer lower-mass BBH mergers than predicted by cluster formation histories that follow the star formation rates and allow for the abundant formation of massive, metal-rich clusters at low redshift and 2) other dynamical environments such as nuclear star clusters and the disks of active galactic nuclei \citep[e.g.,][]{Antonini_Rasio_2016, Gondan+2018,Fragione+2019,Tagawa+2020} could supply BBH mergers with $M_p\gtrsim 50\,$\msun\ in addition to globular clusters and young massive star clusters.

\section{Conclusions}\label{sec:conclu}
We demonstrated the merger rate distribution as a function of primary mass in the local Universe for BBH mergers from dense star clusters (cluster mass $\gtrsim 10^5$~\msun\, and core density $\gtrsim 10^3\,M_{\odot}\,{\rm pc^{-3}}$). This distribution agrees with the overall BBH merger rate distribution detected by LVK, where about $90\%$ of the mergers have primary mass $M_p<20\,$\msun. Most of these low-mass mergers are produced by dense star clusters with metallicity around $Z_{\odot}$ and contain only first-generation BHs. On the other hand, more massive BBH mergers are predominantly produced in metal-poor globular clusters, and the majority contain higher-generation BHs--particularly for the most massive systems with $M_p>50\,$\msun. Although they comprise a small fraction of low-mass mergers, most hierarchical mergers have primary masses $\lesssim20$~\msun. These findings suggest that the BH spin distributions for low-mass and high-mass cluster BBH mergers may differ, with a larger fraction of high-mass mergers having spins $\gtrsim 0.4$ \citep{Antonini+2025}. We suggested that a significant fraction of low-mass BBH mergers detected by LVK may have isotropic spin distribution as expected from the dynamical formation channel in dense star clusters. This is supported by previous studies, including \citet{Fishbach_Fragione_2023}, which found that BBHs consistent with having isotropic spin directions could account for $61^{+29}_{-44}\%$ of GW detections in the Third GW Transient Catalog, and \citet{Tong+2022}, which found that up to $89\%$ of BBHs mergers may be dynamically formed. GW detections of low-mass BBH mergers can thus provide constraints on the formation of metal-rich dense star clusters at low redshift. 

\begin{acknowledgments}
 We thank Aditya Vijaykumar and Carl Rodriguez for helpful discussions and the anonymous referee for helpful comments. C.S.Y. and M.R.C. acknowledge support from the Natural Sciences and Engineering Research Council of Canada (NSERC) DIS-2022-568580. M.F. is supported by NSERC RGPIN-2023-05511, the University of Toronto Connaught Fund, and the Alfred P. Sloan Foundation.

The BBH merger data from the \texttt{CMC} catalog models used in this study can also be found at https://doi.org/10.5281/zenodo.15832905.
\end{acknowledgments}

\vspace{5mm}

\bibliography{BBH_Mass}
\bibliographystyle{aasjournal}

\end{document}